\documentclass{article}

\usepackage{PRIMEarxiv}

\usepackage[utf8]{inputenc} 
\usepackage[T1]{fontenc}    
\usepackage{hyperref}       
\usepackage{url}            
\usepackage{booktabs}       
\usepackage{amsfonts}       
\usepackage{nicefrac}       
\usepackage{microtype}      
\usepackage{lipsum}
\usepackage{fancyhdr}       
\usepackage{graphicx}       
\graphicspath{{media/}}     

\title{Comprehensive Analysis of BB84, A Quantum Key Distribution Protocol
}

\author{
  SujayKumar Reddy M \\
    School of Computer Science and Engineering  \\
  Vellore Institute of Technology, Vellore\\
  \texttt{sujaykumarreddy.m2020@vitstudent.ac.in} \\
   \And
  Chandra Mohan B \\
  School of Computer Science and Engineering \\
  Vellore Institute of Technology, Vellore\\
  \texttt{chandramohan.b@vit.ac.in} \\
}

\begin{document}
\maketitle

\begin{abstract}
Quantum Key Distribution (QKD) is a technique that enables secure communication between two parties by sharing a secret key. One of the most well-known QKD protocols is the BB84 protocol, proposed by Charles Bennett and Gilles Brassard in 1984. In this protocol, Alice
and Bob use a quantum channel to exchange qubits,
allowing them to generate a shared key that is resistant to
eavesdropping. This paper presents a comparative study of
existing QKD schemes, including the BB84 protocol, and
highlights the advancements made in the BB84 protocol
over the years. The study aims to provide a comprehensive
overview of the different QKD schemes and their strengths
and weaknesses and demonstrate QKD's working principles
through existing simulations and implementations. Through
this study, we show that the BB84 protocol is a highly secure
QKD scheme that has been extensively studied and
implemented in various settings. Furthermore, we discuss
the improvements made to the BB84 protocol to enhance its
security and practicality, including the use of decoy states
and advanced error correction techniques. Overall, this
paper provides a comprehensive analysis of QKD schemes,
focusing on the BB84 protocol in secure communication
technologies.
\end{abstract}


\section{Introduction}
Quantum cryptography stands at the forefront of cryptographic advancements, harnessing the principles of quantum mechanics to establish communication channels with unparalleled security between two entities. While classical cryptosystems have long been employed for secure communication, the escalating sophistication of cyber threats has necessitated the exploration of more robust cryptographic methodologies. The vulnerabilities inherent in classical systems have prompted the evolution of advanced cryptographic techniques, with public key cryptosystems emerging as a notable solution. These systems deploy pairs of keys for encryption and decryption, offering a secure communication framework without the necessity of a shared secret key. However, the susceptibility of public key cryptosystems to hacking attempts has spurred the rise of quantum cryptography.

Quantum cryptography redefines secure communication by leveraging the principles of quantum mechanics to distribute encryption keys. Diverging from classical cryptographic approaches, quantum cryptography derives its security from the intrinsic laws of physics, rendering it exceptionally resistant to even the most sophisticated computing systems. At the core of quantum cryptography lies Quantum Key Distribution (QKD), an innovative protocol that exploits quantum mechanical principles for secure key distribution. The fundamental tenet of QKD revolves around the disturbance caused by the act of measuring a quantum system, thereby enabling the immediate detection of any unauthorized interception of the key. This inherent property preserves the confidentiality and integrity of communication, ensuring that the key distribution process remains secure, with any interception attempts being promptly identified.

As highlighted by Kalaivani et al. \cite{8}, there exists a nuanced distinction between the terms "Quantum Cryptography" and "Quantum Key Distribution" (QKD). Quantum Cryptography, utilizing the Heisenberg Uncertainty Principle and the Principle of Photon Polarization \cite{7}, encapsulates a broader spectrum of quantum-based cryptographic concepts. On the other hand, Quantum Key Distribution specifically employs the principles of Superposition and Quantum Entanglement \cite{8} to establish secure communication channels. This differentiation underscores the diverse and intricate nature of quantum cryptographic techniques, each playing a vital role in advancing the security landscape of modern communication systems. Furthermore, this paper delves into the evolution of the BB84 protocol, elucidating the enhancements devised to bolster its security and practicality. Noteworthy improvements discussed herein include the incorporation of decoy states and advanced error correction techniques. These refinements contribute to fortifying the BB84 protocol against potential vulnerabilities, rendering it not only secure but also practical for real-world applications. In essence, our study provides a thorough and insightful analysis of QKD schemes, with a particular spotlight on the BB84 protocol, contributing to the understanding and advancement of secure communication technologies.

\subsection{Heisenberg Uncertainity Principle}
The Heisenberg uncertainty principle is a foundational
principle of quantum mechanics that arises from the
wave-particle duality of quantum objects. It states that
certain pairs of physical properties of a particle or system
cannot be precisely measured simultaneously. This
principle has important implications for quantum
cryptography, which relies on the fact that any attempt to
measure or intercept a quantum system carrying a secret
key will inevitably disturb its state in a detectable way \cite{9}.
Eavesdropping on a quantum message is therefore akin to
making a measurement, and the uncertainty principle
ensures that any attempt to do so will leave a detectable
signature, allowing legitimate parties to detect and
prevent eavesdropping.

\subsection{Quantum Superposition}
Quantum superposition \cite{1} is a fundamental principle of
quantum mechanics that has been extensively studied and
applied in various fields such as quantum computing and
quantum cryptography. It describes the ability of quantum particles to exist in multiple states simultaneously, and this
property has been observed experimentally in numerous
systems, ranging from photons to atoms and molecules.
One example of quantum superposition is the famous
Schrödinger's cat experiment, which has been extensively
discussed in the literature. Gerry et al and Knight et al \cite{2}
provided a detailed mathematical interpretation of this
experiment, showing that the cat is in a superposition state
of both dead and alive until it is observed, at which point
the wave function collapses into one of the two possible
states.

\section{Quantum Entaglement}

In the realm of quantum mechanics, the phenomenon of
quantum entanglement is observed when two or more
particles become so strongly correlated that their states
are interdependent, regardless of the distance between
them. This phenomenon was referred to as "Spooky Action
of Distance" by Einstein \cite{4}. The unique nature of quantum
entanglement ensures that no other system can be
correlated to the entangled particles, making it a
promising tool for secure communication \cite{11}. The use of
entanglement in communication protocols provides a
means of detecting any attempts at eavesdropping, since
any such attempt would result in a change to the
entangled state that would be detectable to the
communicating parties \cite{10}.

Yixuan et al \cite{3} provide a comprehensive overview of
the process of generating entangled photon pairs using the
Spontaneous Parametric Down Conversion (SPDC)
technique. They discuss the Highly Efficient Source of
Photon Pairs based on SPDC in a Single-Mode Fiber (HSPS)
method, which is often utilized as the source of photons in
Quantum Key Distribution (QKD) systems. The entangled
photons generated via the HSPS method allow for the
creation of a shared secret key that is inherently secure.
Any attempt to eavesdrop or intercept the photons would
cause the entanglement to be destroyed, alerting the
communicating parties to the presence of an attacker.

This comprehensive paper unfolds across five meticulously crafted sections, each contributing to the nuanced exploration of Quantum Key Distribution (QKD) and, notably, the BB84 protocol. Sections 2 and 3 delve into the foundational principles of the QKD protocol, emphasizing a detailed analysis of the renowned BB84 variant. Shifting focus, Sections 4 and 5 undertake a critical examination of diverse iterations of the BB84 protocol crafted by different authors, offering a thorough review of its nuanced variations. The study reaches its zenith in Section 6, where various simulations are executed to empirically scrutinize both the Quantum Key Distribution Protocol and the BB84 variant, providing valuable insights into their real-world applicability and performance under varying conditions.

\section{Quantum Key Distribution (QKD)}
Quantum Key Distribution (QKD) is a cryptographic
technique that allows two parties to distribute a secret key
using quantum mechanics principles. QKD protocols, such
as BB84, decoy-based QKD, and phase differential shift
QKD, have been developed to achieve secure key
distribution. By utilizing quantum properties such as the
uncertainty principle, QKD enables the generation of a
shared secret key that cannot be intercepted without being
detected, providing a high level of security for
cryptographic communications.

\section{BB84 Protocol}

The BB84 protocol, devised by Bennett and Brassard in
1984, is a quantum key distribution (QKD) protocol that
utilizes a combination of a quantum channel and an
insecure classical channel which needs to be authenticated
inorder to distribute a secret key between two parties. The
protocol ensures the security of the key by detecting any
attempts at eavesdropping on the quantum channel, thus
providing a means of detecting any potential security
breaches.
The BB84 protocol uses four quantum states that are
randomly prepared by the sender, Alice, in one of two
bases (rectilinear basis or diagonal basis) to transmit
information to the receiver, Bob.

The four quantum states used in BB84 are:
\begin{enumerate}
    \item $|0\rangle$ and $|1\rangle$ in the rectilinear basis (X-basis), where
    $|0\rangle = (1, 0)$ and $|1\rangle = (0, 1)$
    \item $|+\rangle$ and $|-\rangle$ in the diagonal basis (Z-basis), where
    $|+\rangle = \frac{1}{\sqrt{2}}(1, 1)$ and $|-\rangle = \frac{1}{\sqrt{2}}(1, -1)$
\end{enumerate}
The working principle of BB84 protocol is stated below
with Alice and Bob.

\begin{figure}
  \centering
  \includegraphics[scale=0.2]{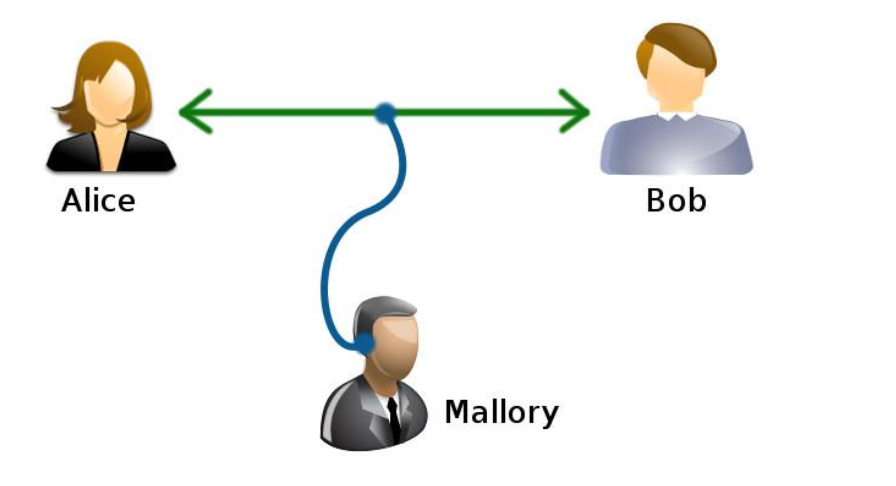}
  \caption{Alice, Bob and Mallory Classic Communication Example}
  \label{fig:fig1}
\end{figure}

\begin{figure}
  \centering
  \includegraphics[scale=0.2]{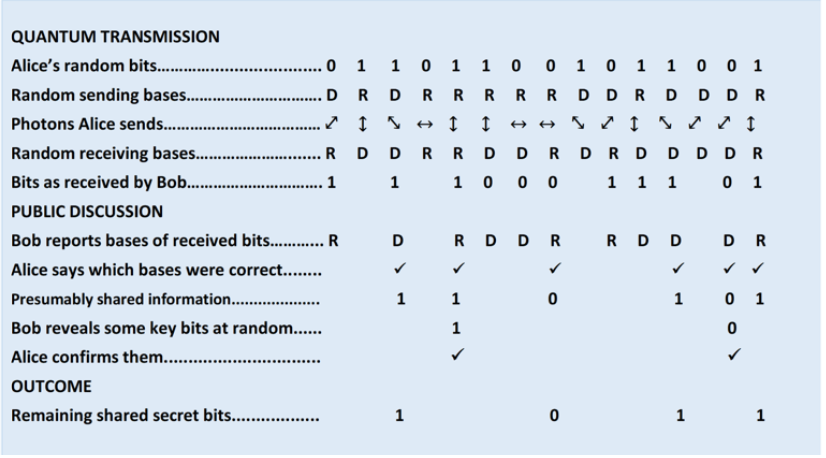}
  \caption{Transmission from Alice to Bob without Mallory in the Quantum Channel using BB84 Protocol}
  \label{fig:fig1}
\end{figure}

\begin{figure}
  \centering
  \includegraphics[scale=0.2]{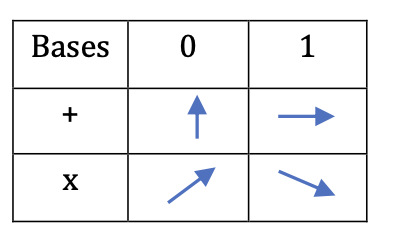}
  \caption{Rectilinear and Diagonal Bases according to the
Quantum Transmission}
  \label{fig:fig1}
\end{figure}

In Figure 1, we can see the communication process between Alice and Bob, with Mallory being the potential intruder who may attempt to intercept their communication. In Figure 2, as presented in \cite{16}, the transmission of bits between Alice and Bob can be observed, which ultimately leads to the generation of the Shared Secret Key, also known as the sifted key in Quantum Key Distribution (QKD). The BB84 protocol consists of the following steps:
\begin{enumerate}
    \item Alice generates a random sequence of bits and encodes each bit into a randomly chosen polarization state using one of two bases: Rectilinear (R) or Diagonal (D).
    \item However, when information is encoded in non-orthogonal quantum states, such as single photons with polarization directions 0, 45, 90, and 135 degrees according to Figure \ref{fig:figure3}.
    \item Alice sends the encoded photons to Bob through the quantum channel.
    \item Bob randomly chooses one of the two bases to measure the polarization of each photon he receives from Alice.
    \item Bob measures each photon and records the results as either a 0 or a 1.
    \item After the transmission, Alice and Bob publicly disclose the bases they used for each bit over the classical channel.
    \item Alice and Bob discard the bits where the bases did not match.
    \item The remaining bits form the sifted key.
    \item To ensure the security of the key, Alice and Bob perform a test to detect any potential eavesdropping by comparing a subset of their sifted key over the classical channel.
    \item Finally, they use error correction and privacy amplification techniques to extract a shorter, but secure, shared secret key.
\end{enumerate}

To ensure the utmost security of the key generated through the BB84 protocol, Alice and Bob meticulously conduct a pivotal test designed to detect potential eavesdropping. This involves a thoughtful comparison of a randomly chosen subset of their sifted key over the classical channel. Alice communicates her chosen subset to Bob, who then scrutinizes it against his own subset, promptly sharing the results with Alice. A substantial correlation between the subsets signifies the absence of an eavesdropper on the quantum channel, affirming the key's security. Conversely, a lack of correspondence indicates potential eavesdropping, prompting an immediate protocol termination. Termed the "privacy amplification" step, this test is integral in safeguarding key distribution. Concluding the protocol, error correction and privacy amplification techniques are employed to derive a shorter yet highly secure shared secret key. The accuracy of the key's authenticity depends on whether Alice and Bob measured in the same bases, resulting in 100\% accuracy, or if different bases were chosen, reducing accuracy to 50\%. In essence, the BB84 protocol ingeniously enables two parties to establish a shared secret key over an insecure channel, harmonizing the principles of quantum mechanics with classical communication.

\section{Related Work}
The study done by Meyer et al and his colleagues \cite{12}
conducted a survey of the current state of research in
quantum cryptography, with a specific focus on the BB84
protocol and the quantum bit commitment (QBC) protocol.
The authors provided a detailed explanation of the working
principle of the BB84 protocol and presented proofs for
three concepts of "unconditional security" in quantum key
distribution. These three concepts are the entanglement-
based version of the protocol, the equivalent
entanglement-based version, and a version where these
two versions are shown to be equivalent. The QBC protocol
was identified as an active area of research within quantum
cryptography.
Sasirekha et al and Hemalatha et al \cite{6} have conducted
extensive research on Quantum Key Distribution (QKD)
and its practical applications. Their work includes a
detailed outline of the six critical steps involved in QKD, as
well as proposed methods for detecting eavesdropping.
These methods include measuring the polarization of
photons, which cannot be done without disturbing the
photon, and using entangled photons, which allows Alice
and Bob to detect any interference or disturbance in the protocol. The researchers also highlight the potential
applications of QKD, including IoT and secure voting, as
demonstrated by a case study in Switzerland \cite{13}.
Abushgra et al and Abdulbast et al conducted an extensive
investigation \cite{14} into multiple Quantum Key Distribution
(QKD) algorithms, which included a comparison of the
performance of the BB84 protocol with its various
iterations, such as B92, SARG04 \cite{15}, KMB09, EPR, S13, and
DPS. The study focused on the working principles of these
algorithms and evaluated their effectiveness in detecting
presence, polarization, state probability, qubit string,
classical channel presence, decoy states, sifting phase, and
different attacks in QKD. The results of the study showed
that the BB84 protocol was vulnerable to some attacks.

Priyanka et al. \cite{16} identified various limitations of the
BB84 protocol, including the introduction of noise caused
by imperfect detectors, technical difficulties in producing a
perfectly single-photon pulse, inefficiency in generating the
key, and the assumption that the subset of the total
transmission will be used to generate the key. The paper
also delves into the modifications made to the BB84
protocol, such as omitting the public announcement of
bases, using multilevel encoding, implementing the decoy
pulse method, and employing bi-directional QKD with
practical faint pulses.

\section{Recent Enhancements in BB84}
In the study \cite{17}, the authors focused on improving the
BB84 protocol using the polarization technique, which
allows for the transmission of n-bit keys via photons
without any loss of information. They proposed an
enhanced BB84 protocol (EBB84) that eliminates the need
for a classical channel by allowing the sender to randomly
encode n-bit keys into photons and transmit them through
a quantum channel. The receiver then generates the key
using the same polarization, and both parties agree on the
basis used for encoding and transmitting the key. The
authors used simulations to compare the performance of
the BB84 and EBB84 protocols in terms of execution time,
key length parameter, and Quantum Bit Error Rate (QBER).
The simulations were carried out on a computer with
specific hardware and software specifications, and the
results were presented in tables and figures.

In the study \cite{18}, the authors propose a modified version of
the BB84 protocol to overcome security issues caused by
various source imperfections, including state preparation
flaws and side channels such as Trojan-horse attacks, mode
dependencies, and classical correlations between emitted
pulses. The proposed protocol exploits basis mismatched
events and employs the Reference Technique, a powerful
mathematical tool, to ensure implementation security. The
authors also compare the achievable secret-key rate of the
modified BB84 protocol with that of the three-state loss-tolerant protocol and demonstrate that adding a fourth
state significantly improves the estimation of leaked
information in the presence of source imperfections,
resulting in better performance.

In their research \cite{20}, the authors introduce the BB84
protocol as the first quantum confidential communication
protocol and propose a QKD network structure specifically
designed for power business scenarios, which accounts for
the complexity and diversity of power grid environments
and communication transmission losses. To evaluate the
performance of the QKD system, the study analyzes two
tiers - the key tier, which focuses on factors that may affect
the secret key rate in the quantum channel, and the
business tier, which examines the transmission
performance of the system when using a quantum virtual
private network (QVPN) for encrypted transmission. The
performance of the QKD device and QVPN is then tested in
various simulated environments, providing insights into
the system's suitability for large-scale applications.

The paper \cite{21} delves into the security concerns and risks
that arise when transmitting sensitive medical data
through wireless body sensor networks (WBSNs). To
mitigate these vulnerabilities, the paper proposes an
enhanced BB84 quantum cryptography protocol
(EBB84QCP) that uses a bitwise operator for secure secret
key sharing between communicating parties. This
approach avoids the direct sharing of keys through
traditional methods like email or phone, which are
susceptible to interception by attackers. The proposed
protocol is demonstrated to be efficient in terms of the
constrained resources of WBSNs and provides a robust
level of security for transmitting sensitive information.

\section{Simulation Results of Quantum Key Distribution , BB84 Protocol}

\subsection{BB84 Alice- Bob Simulation Results \cite{23}}
The paper \cite{22} proposes some existing simulations and we
choose do simulations with this tool for different
parameters.

\begin{figure}
  \centering
  \includegraphics[scale=0.6]{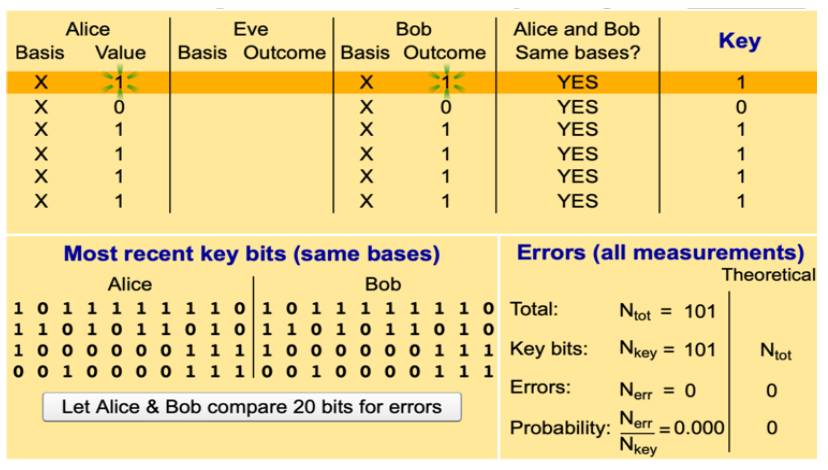}
  \caption{Alice and Bob have same bases(X)}
  \label{fig:fig1}
\end{figure}

\begin{figure}
  \centering
  \includegraphics[scale=0.6]{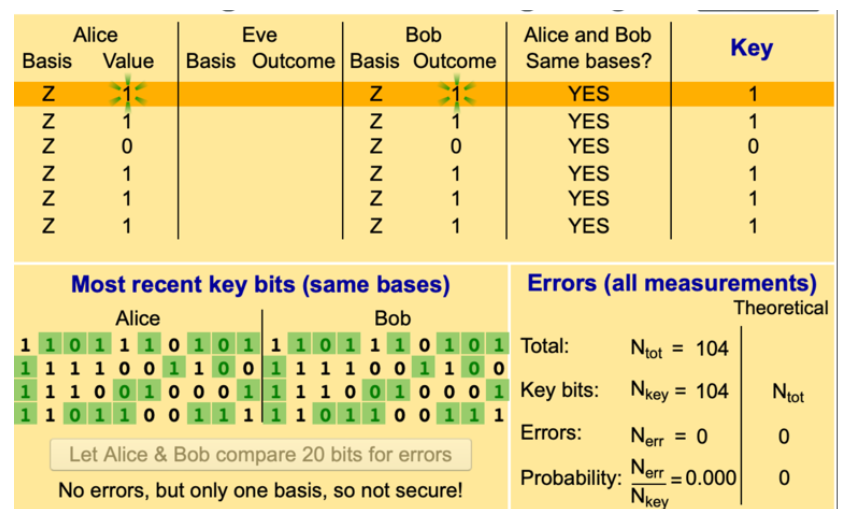}
  \caption{Alice and Bob have same bases(Z)}
  \label{fig:fig1}
\end{figure}

\begin{figure}
  \centering
  \includegraphics[scale=0.6]{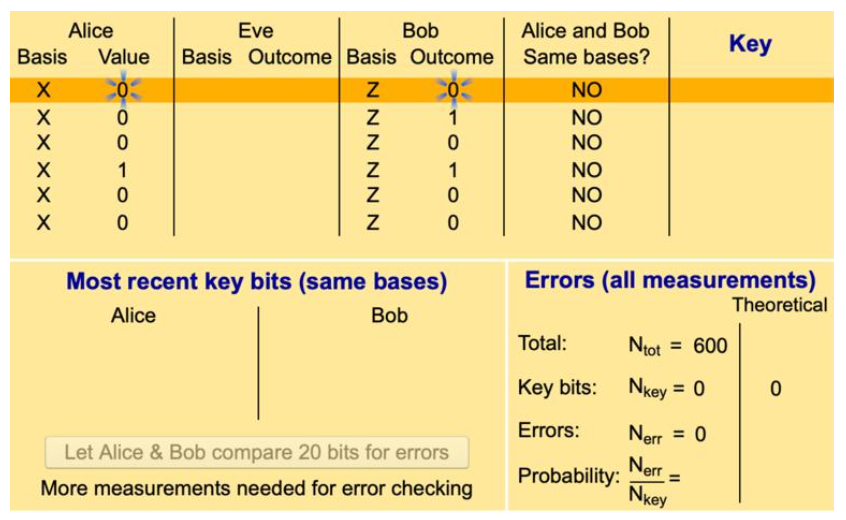}
  \caption{Alice(X) and Bob(Z) have different bases}
  \label{fig:fig1}
\end{figure}

\begin{figure}
  \centering
  \includegraphics[scale=0.6]{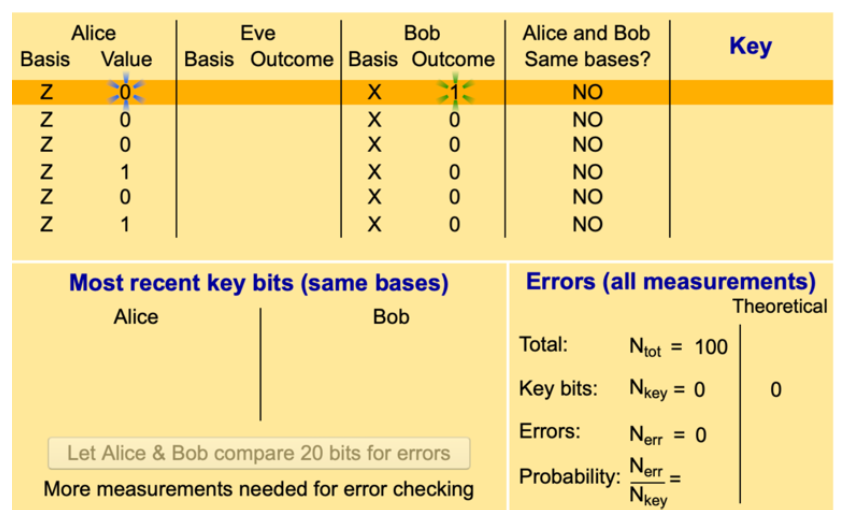}
  \caption{Alice(Z) and Bob(X) have different bases}
  \label{fig:fig1}
\end{figure}

\begin{figure}
  \centering
  \includegraphics[scale=0.6]{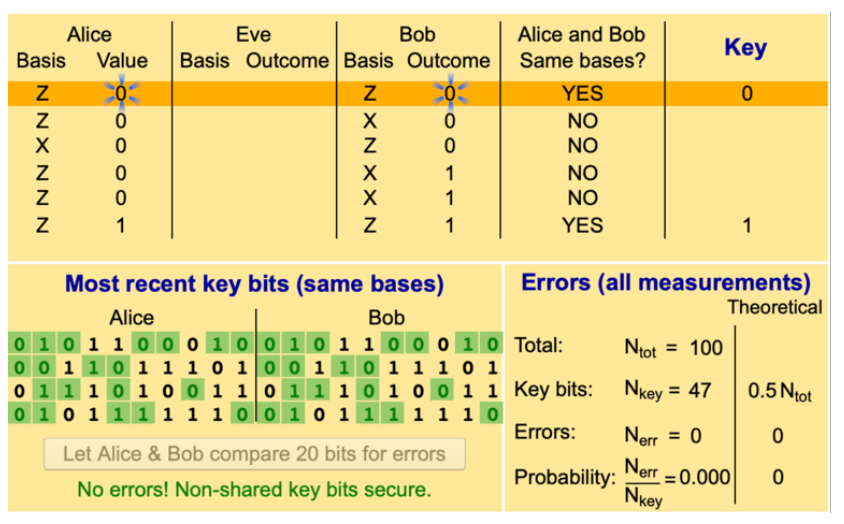}
  \caption{Alice and Bob have random orientation}
  \label{fig:fig1}
\end{figure}

\begin{figure}
  \centering
  \includegraphics[scale=0.6]{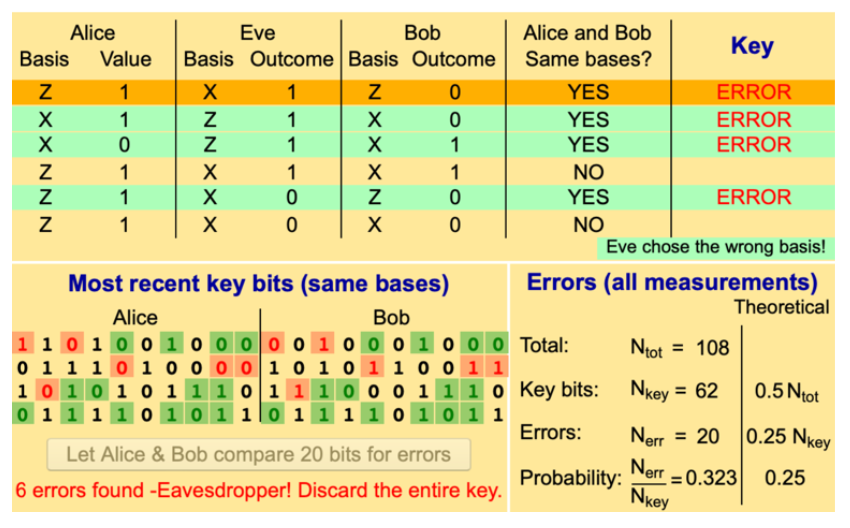}
  \caption{Alice and Bob have random orientation with Eve
intercepting and resend particles}
  \label{fig:fig1}
\end{figure}

\begin{table}[ht]
    \centering
    \caption{Nodes Location for only Key Generator Layer}
    \label{table:node-locations}
    \begin{tabular}{c|c|c}
        \hline
        \textbf{Node s.no} & \textbf{LOCATION} & \textbf{LAYER} \\
        \hline
        Node 1 & MUMBAI & QKD-KMS Default \\
        Node 2 & DELHI & QKD-KMS Default \\
        Node 3 & NEAR TO MUMBAI & 3-4 Key Generator Layer \\
        Node 4 & NEAR TO DELHI & 3-4 Key Generator Layer \\
        \hline
    \end{tabular}
\end{table}
\begin{table}[ht]
    \centering
    \caption{Nodes Parameters for only Key Generator Layer}
    \label{table:node-parameters}
    \begin{tabular}{|c|c|c|}
        \hline
        \textbf{Layer} & \textbf{Nodes} & \textbf{Parameters} \\
        \hline
        1-2 & QKD KMS & Node 1 and Node 2 \\
        & & Key Size (BIT) – 512 \\
        & & PP Packet size – 100 \\
        & & OKD Stop Time – 50sec \\
        \hline
        3-4 & Key Generator Layer & Node 3 and Node 4 \\
        & & Average size of consumed key-pairs (bits) - 0 \\
        & & Average size of generated key-pairs (bits) - 512 \\
        & & Key rate (bit/sec) - 10811 \\
        & & Key-pairs consumed - 0 \\
        & & Key-pairs consumed (bits) - 0 \\
        & & Key-pairs generated - 1055 \\
        & & Key-pairs generated (bits) - 540160 \\
        & & Link distance (meters) - 1441565 \\
        & & QKD Buffer Capacity (bits) - 50000000 \\
        & & Start Time (sec) - 0 \\
        & & Stop Time (sec) - 50 \\
        \hline
    \end{tabular}
\end{table}
\begin{figure}
  \centering
  \includegraphics[scale=0.6]{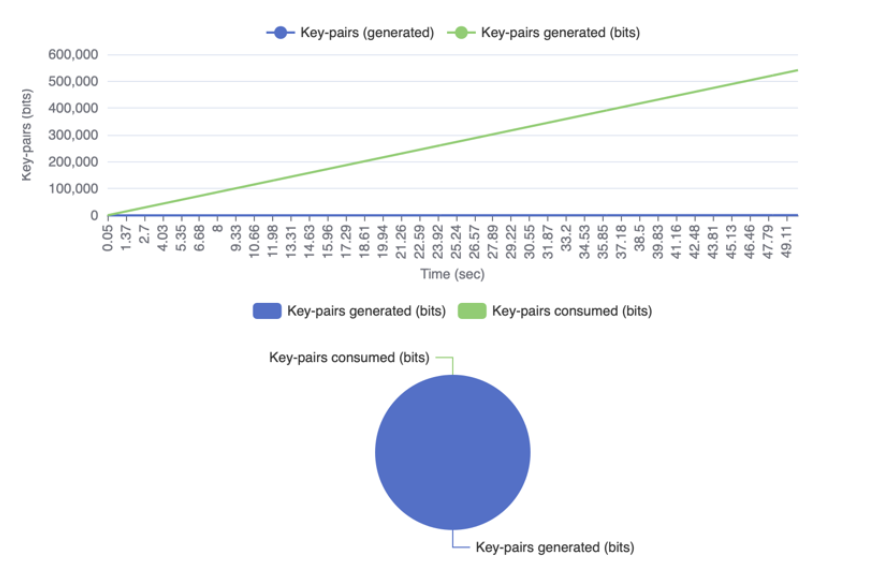}
  \caption{Key Generated Pairs for Key Generated Layer}
  \label{fig:fig1}
\end{figure}

\begin{table}[ht]
    \centering
    \caption{Nodes Location for ETSI 014 and Key Generator Layer}
    \label{table:etsi014-node-locations}
    \begin{tabular}{|c|c|c|}
        \hline
        \textbf{Node s.no} & \textbf{LOCATION} & \textbf{LAYER} \\
        \hline
        Node 1 & MUMBAI & QKD-KMS Default \\
        Node 2 & DELHI & QKD-KMS Default \\
        Node 3 & NEAR TO MUMBAI & 3-4 Key Generator Layer \\
        Node 4 & NEAR TO DELHI & 3-4 Key Generator Layer \\
        Node 5 & PUNE & 5-6 Key Consumer Layer (ETSI 014) \\
        Node 6 & GWALIOR & 5-6 Key Consumer Layer (ETSI 014) \\
        \hline
    \end{tabular}
\end{table}
\begin{table}[ht]
    \centering
    \caption{Nodes Parameters for ETSI 014 and Key Generator Layer}
    \label{table:etsi014-node-parameters}
    \begin{tabular}{|c|c|c|}
        \hline
        \textbf{Layer} & \textbf{Nodes} & \textbf{Parameters} \\
        \hline
        1-2 & QKD KMS & Node 1 and Node 2 \\
        & & Key Size (BIT) – 512 \\
        & & PP Packet size – 100 \\
        & & OKD Stop Time – 50sec \\
        \hline
        3-4 & Key Generator Layer & Node 3 and Node 4 \\
        & & Average size of consumed key-pairs (bits) - 487 \\
        & & Average size of generated key-pairs (bits) - 498 \\
        & & Key rate (bit/sec) - 100000 \\
        & & Key-pairs consumed - 2938 \\
        & & Key-pairs consumed (bits) - 1432256 \\
        & & Key-pairs generated - 11517 \\
        & & Key-pairs generated (bits) - 5735936 \\
        & & Link distance (meters) - 1410425 \\
        & & QKD Buffer Capacity (bits) - 50000000 \\
        & & Start Time (sec) - 0 \\
        & & Stop Time (sec) - 50 \\
        \hline
        5-6 & Key Consumer Layer ETSI 014 & Node 5 and Node 6 \\
        & & Key Consumption Statistics \\
        & & Average size of consumed key-pairs (bits) - 464 \\
        & & Key-pairs consumed - 1500 \\
        & & Key-pairs consumed (bits) - 696000 \\
        & & QKD Apps Statistics \\
        & & Authentication - SHA-1 \\
        & & Bytes Received - 154294 \\
        & & Bytes Sent - 154294 \\
        & & Encryption - OTP \\
        & & Key/Data utilization (\%) - 74.91 \\
        & & Missed send packet calls - 251 \\
        & & Number of Keys to Fetch From KMS - 3 \\
        & & Packet Size (bytes) - 100 \\
        & & Packets Received - 749 \\
        & & Packets Sent - 749 \\
        & & Start Time (sec) - 10 \\
        & & Stop Time (sec) - 50 \\
        & & Traffic Rate (bit/sec) - 20000 \\
        & & QKD Apps-KMS Statistics \\
        & & Bytes Received - 978368 \\
        & & Bytes Sent - 572684 \\
        & & Packets Received - 1439 \\
        & & Packets Sent - 1439 \\
        & & Signaling Statistics \\
        & & Bytes Received - 365000 \\
        & & Bytes Sent - 365000 \\
        & & Packets Received - 1000 \\
        & & Packets Sent - 1000 \\
        \hline
    \end{tabular}
\end{table}
\begin{figure}
  \centering
  \includegraphics[scale=0.6]{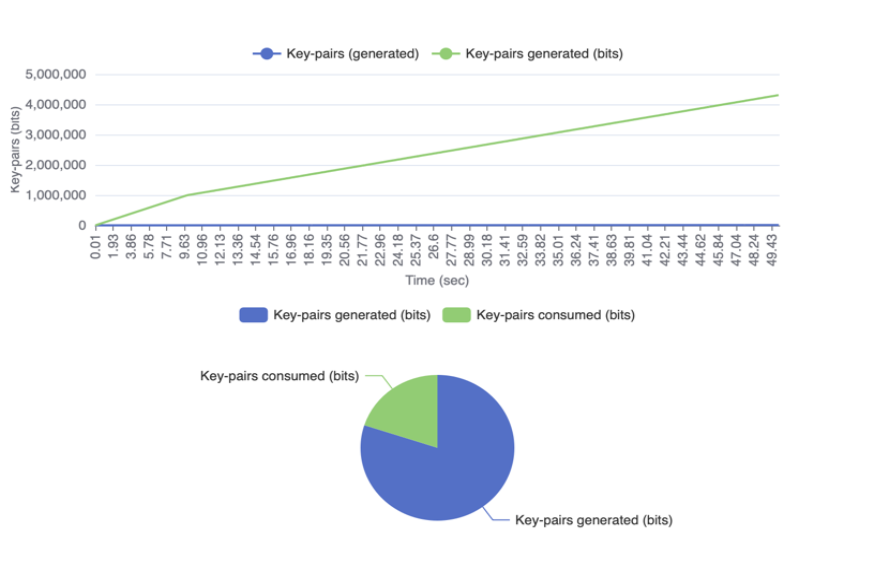}
  \caption{Key Generated Pairs for ETSI 014 and Key
Generator Layer}
  \label{fig:fig1}
\end{figure}
\begin{figure}
  \centering
  \includegraphics[scale=0.6]{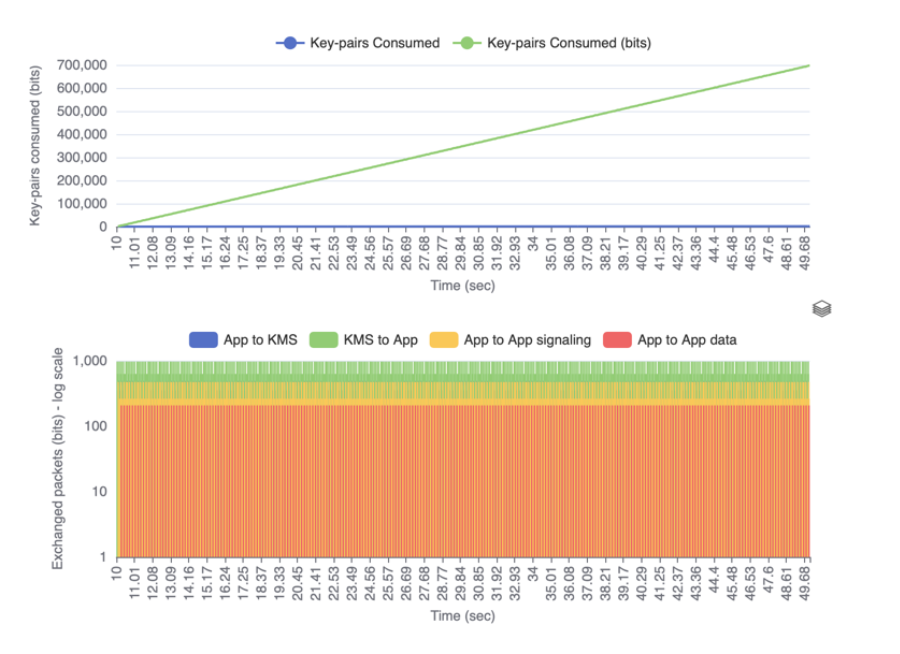}
  \caption{Key Consumer Layer ETSI 014 Statistics}
  \label{fig:fig1}
\end{figure}

\begin{table}[ht]
    \centering
    \caption{Nodes Location for ETSI 004 and Key Generator Layer}
    \label{table:etsi004-node-location}
    \begin{tabular}{|c|c|c|}
        \hline
        \textbf{Node s.no} & \textbf{LOCATION} & \textbf{LAYER} \\
        \hline
        Node 1 & MUMBAI & QKD-KMS Default \\
        Node 2 & DELHI & QKD-KMS Default \\
        Node 3 & NEAR TO MUMBAI & 3-4 Key Generator Layer \\
        Node 4 & NEAR TO DELHI & 3-4 Key Generator Layer \\
        Node 5 & PUNE & 5-6 Key Consumer Layer (ETSI 004) \\
        Node 6 & GWALIOR & 5-6 Key Consumer Layer (ETSI 004) \\
        \hline
    \end{tabular}
\end{table}
\begin{table}[ht]
    \centering
    \caption{Nodes Parameters for ETSI 004 and Key Generator Layer}
    \label{table:etsi004-node-parameters}
    \begin{tabular}{|c|c|c|}
        \hline
        \textbf{Layer} & \textbf{Nodes} & \textbf{Parameters} \\
        \hline
        1-2 & QKD KMS & Node 1 and Node 2 \\
        & & Key Size (BIT) – 512 \\
        & & PP Packet size – 100 \\
        & & OKD Stop Time – 50sec \\
        \hline
        3-4 & Key Generator Layer & Node 3 and Node 4 \\
        & & Average size of consumed key-pairs (bits) - 512 \\
        & & Average size of generated key-pairs (bits) - 512 \\
        & & Key rate (bit/sec) - 100000 \\
        & & Key-pairs consumed - 1796 \\
        & & Key-pairs consumed (bits) - 919552 \\
        & & Key-pairs generated - 9767 \\
        & & Key-pairs generated (bits) - 5003456 \\
        & & Link distance (meters) - 1440197 \\
        & & QKD Buffer Capacity (bits) - 50000000 \\
        & & Start Time (sec) - 0 \\
        & & Stop Time (sec) - 50 \\
        \hline
        5-6 & Key Consumer Layer ETSI 004 & Node 5 and Node 6 \\
        & & Key Consumption Statistics \\
        & & Average size of consumed key-pairs (bits) - 462 \\
        & & Key-pairs consumed - 1976 \\
        & & Key-pairs consumed (bits) - 914848 \\
        & & QKD Apps Statistics \\
        & & Authentication - SHA-1 \\
        & & Bytes Received - 202910 \\
        & & Bytes Sent - 202910 \\
        & & Encryption - OTP \\
        & & Key/Data utilization (\%) - 100 \\
        & & Missed send packet calls - 0 \\
        & & Packet Size (bytes) - 100 \\
        & & Packets Received - 985 \\
        & & Packets Sent - 985 \\
        & & Size of Key Buffer for Authentication - 6 \\
        & & Size of Key Buffer for Encryption - 1 \\
        & & Start Time (sec) - 10 \\
        & & Stop Time (sec) - 50 \\
        & & Traffic Rate (bit/sec) - 20000 \\
        & & QKD Apps-KMS Statistics \\
        & & Bytes Received - 1978801 \\
        & & Bytes Sent - 1716638 \\
        & & Packets Received - 3964 \\
        & & Packets Sent - 3966 \\
        & & Signaling Statistics \\
        & & Bytes Received - 1812 \\
        & & Bytes Sent - 1812 \\
        & & Packets Received - 6 \\
        & & Packets Sent - 6 \\
        \hline
    \end{tabular}
\end{table}
\begin{figure}
  \centering
  \includegraphics[scale=0.6]{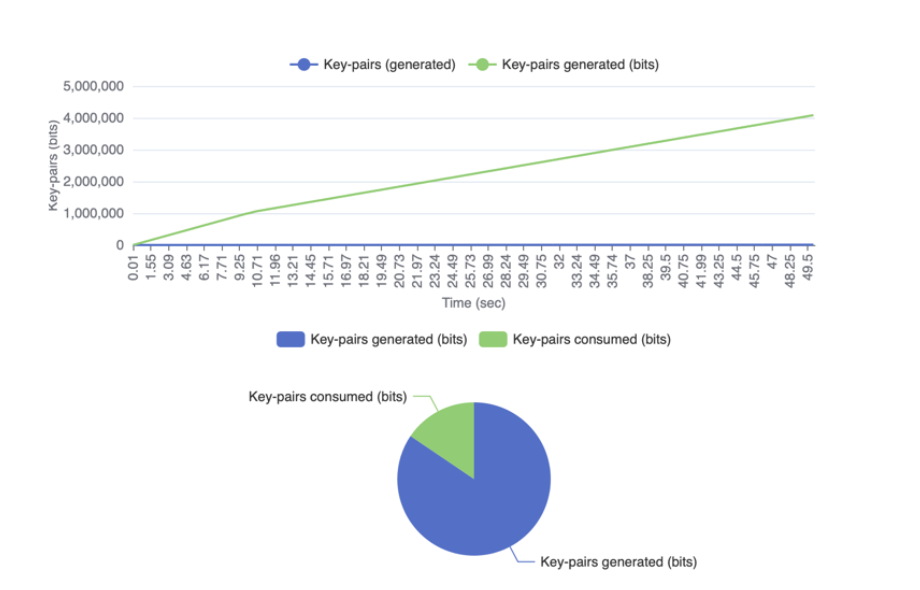}
  \caption{Key Generator Layer for ETSI004 and Key
Generator Layer}
  \label{fig:fig1}
\end{figure}
\begin{figure}
  \centering
  \includegraphics[scale=0.6]{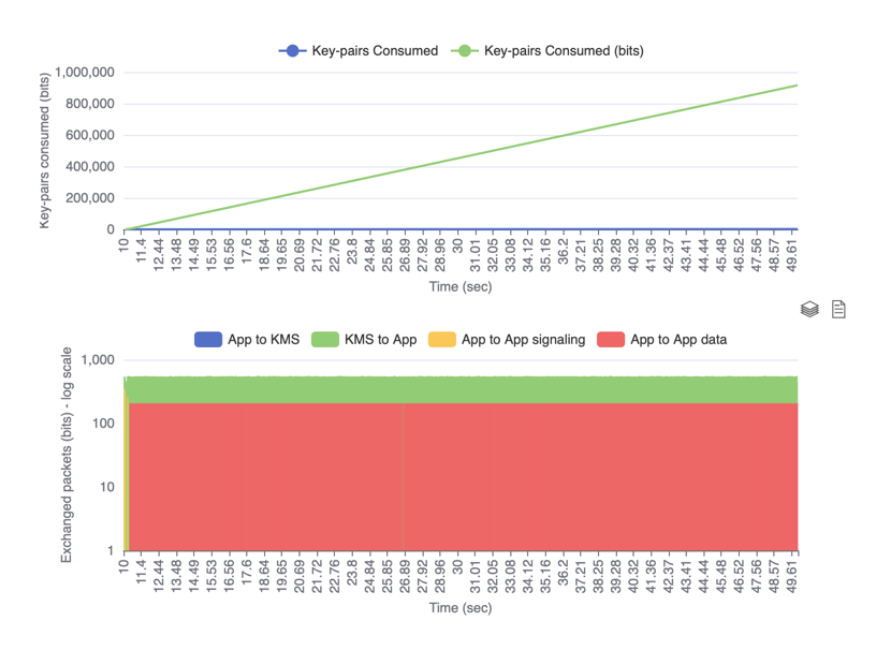}
  \caption{Key Consumer Layer ETSI004 Statistics}
  \label{fig:fig1}
\end{figure}

\begin{table}[ht]
    \centering
    \caption{Nodes Location for ETSI 004, ETSI 014, and Key Generator Layer}
    \label{table:etsi004-etsi014-node-location}
    \begin{tabular}{|c|c|c|}
        \hline
        \textbf{Node s.no} & \textbf{LOCATION} & \textbf{LAYER} \\
        \hline
        Node 1 & MUMBAI & QKD-KMS Default \\
        Node 2 & DELHI & QKD-KMS Default \\
        Node 3 & NEAR TO MUMBAI & 3-4 Key Generator Layer \\
        Node 4 & NEAR TO DELHI & 3-4 Key Generator Layer \\
        Node 5 & PUNE & 5-6 Key Consumer Layer (ETSI 004) \\
        Node 6 & GWALIOR & 5-6 Key Consumer Layer (ETSI 004) \\
        Node 7 & Goa & 5-6 Key Consumer Layer (ETSI 014) \\
        Node 8 & Kolkata & 5-6 Key Consumer Layer (ETSI 014) \\
        \hline
    \end{tabular}
\end{table}
\begin{table}[ht]
    \centering
    \caption{Nodes Parameters for ETSI 004, ETSI 014, and Key Generator Layer}
    \label{table:nodes-parameters-etsi-004-014}
    \begin{tabular}{|c|c|c|}
        \hline
        \textbf{Layer} & \textbf{Nodes} & \textbf{Parameters} \\
        \hline
        1-2 & QKD KMS Node 1 and Node 2 & Key Size (BIT) - 512 \\
        & & PP Packet size - 100 \\
        & & OKD Stop Time - 50sec \\
        \hline
        3-4 & Key Generator Layer Node 3 and Node 4 & Average size of consumed key-pairs (bits) - 479 \\
        & & Average size of generated key-pairs (bits) - 498 \\
        & & Key rate (bit/sec) - 100000 \\
        & & Key-pairs consumed - 4902 \\
        & & Key-pairs consumed (bits) - 2351584 \\
        & & Key-pairs generated - 11519 \\
        & & Key-pairs generated (bits) - 5739488 \\
        & & Link distance (meters) - 1440197 \\
        & & QKD Buffer Capacity (bits) - 50000000 \\
        & & Start Time (sec) - 0 \\
        & & Stop Time (sec) - 50 \\
        \hline
        5-6 & Key Consumer Layer ETSI 004 & Node 5 and Node 6 \\
        & & Key-pairs consumed (bits) - 914848 \\
        & & QKDApps Statistics \\
        & & Authentication - SHA-1 \\
        & & Bytes Received - 202910 \\
        & & Bytes Sent - 202910 \\
        & & Encryption - OTP \\
        & & Key/Data utilization (\%) - 100 \\
        & & Missed send packet calls - 0 \\
        & & Packet Size (bytes) - 100 \\
        & & Packets Received - 985 \\
        & & Packets Sent - 985 \\
        & & Size of Key Buffer for Authentication - 6 \\
        & & Size of Key Buffer for Encryption - 1 \\
        & & Start Time (sec) - 10 \\
        & & Stop Time (sec) - 50 \\
        & & Traffic Rate (bit/sec) - 20000 \\
        & & QKDApps-KMS Statistics \\
        & & Bytes Received - 1978469 \\
        & & Bytes Sent - 1716656 \\
        & & Packets Received - 3964 \\
        & & Packets Sent - 3966 \\
        & & Signaling Statistics \\
        & & Bytes Received - 1812 \\
        & & Bytes Sent - 1812 \\
        & & Packets Received - 6 \\
        & & Packets Sent - 6 \\
        \hline

\end{tabular}
\end{table}
\begin{table}[ht]
    \centering
    \caption{Nodes Parameters for ETSI 004, ETSI 014, and Key Generator Layer}
    \label{table:nodes-parameters-etsi-004-014}
    \begin{tabular}{|c|p{0.3\linewidth}|c|}
        \hline
        \textbf{Layer} & \textbf{Nodes} & \textbf{Parameters} \\
        \hline
        7-8 & Key Consumer Layer ETSI 014 Node 7 and Node 8  & Average size of consumed \\
        & & key-pairs (bits) - 464 \\
        & & Key-pairs consumed - 1500 \\
        & & Key-pairs consumed (bits) - 696000 \\
        & & QKDApps Statistics \\
        & & Authentication - SHA-1 \\
        & & Bytes Received - 154294 \\
        & & Bytes Sent - 154294 \\
        & & Encryption - OTP \\
        & & Key/Data utilization (\%) - 74.91 \\
        & & Missed send packet calls - 251 \\
        & & Number of Keys to Fetch \\
        & & From KMS - 3 \\
        & & Packet Size (bytes) - 100 \\
        & & Packets Received - 749 \\
        & & Packets Sent - 749 \\
    & & Start Time (sec) - 10 \\
    & & Stop Time (sec) - 50 \\
    & & Traffic Rate (bit/sec) - 20000 \\
    & & QKDApps-KMS Statistics \\
    & & Bytes Received - 978368 \\
    & & Bytes Sent - 572684 \\
    & & Packets Received - 1439 \\
    & & Packets Sent - 1439 \\
    & & Signaling Statistics \\
    & & Bytes Received - 365000 \\
    & & Bytes Sent - 365000 \\
    & & Packets Received - 1000 \\
    & & Packets Sent - 1000 \\
    \hline
\end{tabular}
\end{table}
\begin{figure}
  \centering
  \includegraphics[scale=0.6]{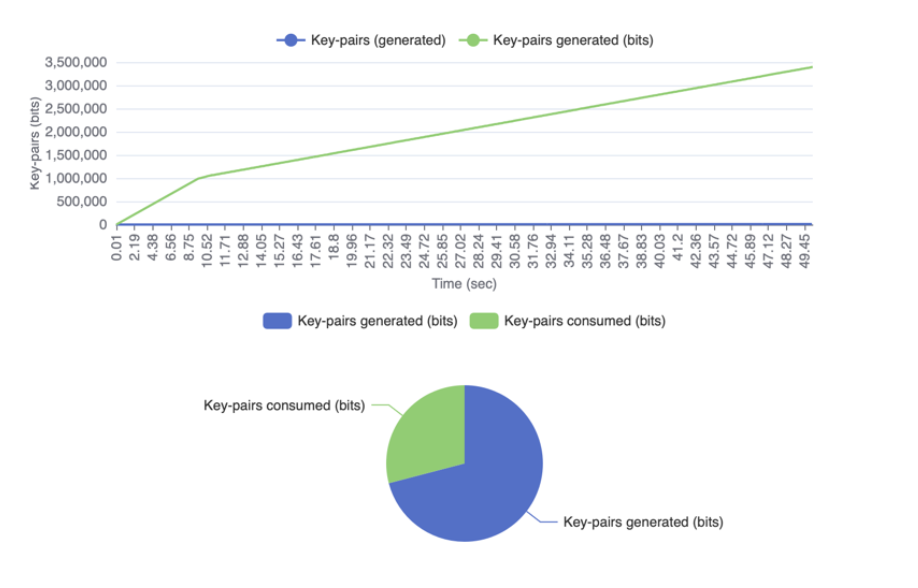}
  \caption{ Key Generator Layer for ETSI 004, ETSI 014
and Key Generator Layer}
  \label{fig:fig1}
\end{figure}
\begin{figure}
  \centering
  \includegraphics[scale=0.6]{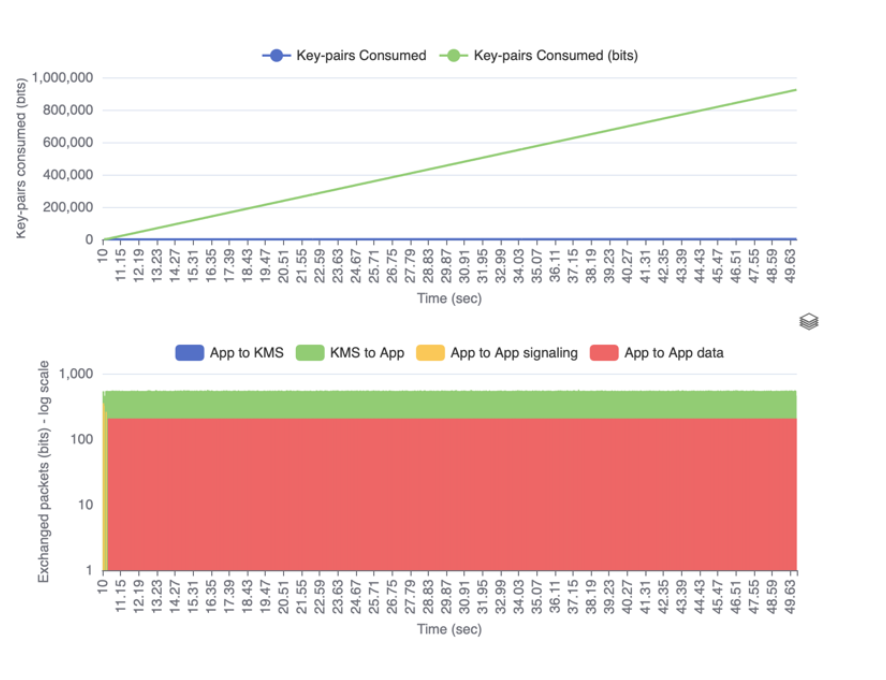}
  \caption{Key Consumer Layer for ETSI 004 Statistics}
  \label{fig:fig1}
\end{figure}
\begin{figure}
  \centering
  \includegraphics[scale=0.6]{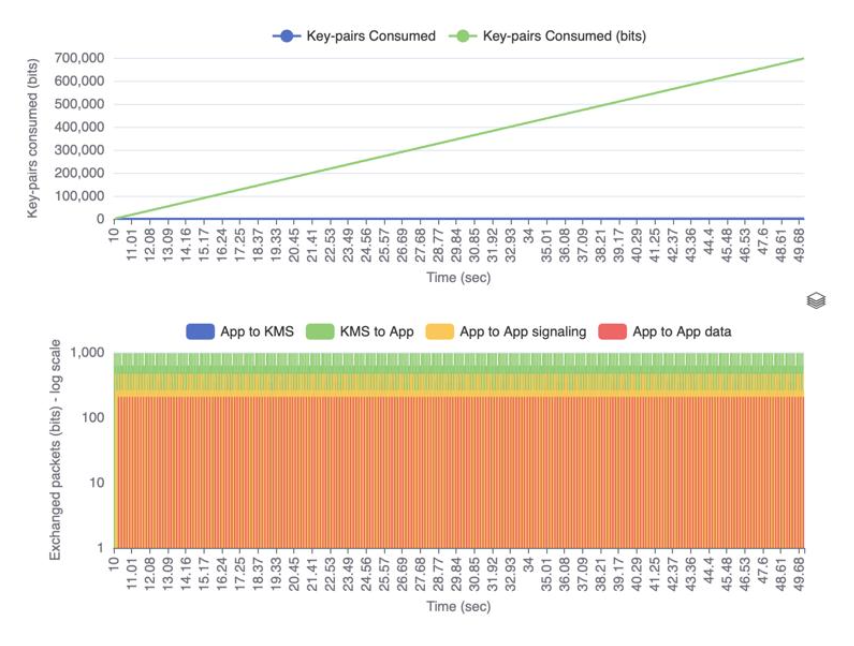}
  \caption{Key Consumer Layer for ETSI 014 Statistics}
  \label{fig:fig1}
\end{figure}

In Figures [4] and [5], the experimental results depict the utilization of X/Z bases by both Alice and Bob, aiming for theoretically zero errors. However, this method is deemed insecure since it involves the use of identical bases by both parties, potentially enabling Eve to intercept the messages by possessing the same bases.

Figures [6] and [7] reveal that no key generation occurs when Alice measures in the X bases and Bob in the Z bases, regardless of the parameters. This observation is based on a fixed orientation approach. To delve deeper into the subject, we further investigate the outcomes of a random orientation approach.

Figure 8 illustrates that employing random bases enhances security between Alice and Bob, as Eve faces the challenge of selecting the same basis as Alice.

The data presented in Figure [9] indicates the occurrence of an error, suggesting a potential interception by Eve. Consequently, it is advisable to discard the entire key to preempt any security breach.

\subsection{Quantum Key Distribution Network Simulation Results \cite{24}}

QKDNetSim serves as a pivotal network simulation platform designed to streamline the integration of Quantum Key Distribution (QKD) into established networks. As QKD research continues to evolve in complexity, the role of simulation technologies has become indispensable for evaluating the practical viability of theoretical advancements. QKDNetSim empowers researchers by providing a platform to construct intricate network topologies and execute repeatable experiments. This capability not only expedites the research process but also mitigates the time and costs typically associated with deploying a comprehensive physical testbed. The platform's versatility contributes to a more efficient and cost-effective exploration of QKD implementations within diverse network environments.

The QKDNetSim is anchored on the QKD-Key Management System (KMS) as the central entity, establishing connections between applications that demand cryptographic operations and the broader QKD network. It adheres to the standards outlined by ETSI004 \cite{26} and ETSI014 \cite{27}. The simulator incorporates various components, including the QKD Key, QKD Buffer, QKD Encryptor, and QKD Post-Processing Application. A distinctive feature of QKDNetSim is its implementation of custom signalization, enabling communication between two KMS. This sets it apart from other simulators \cite{28} and enhances its capabilities in simulating quantum key distribution scenarios.

A notable strength of QKDNetSim lies in its capacity to analyze diverse attack scenarios, including Distributed Denial of Service (DDoS) on Key Management Systems (KMS) \cite{25}. Additionally, QKDNetSim offers the flexibility to integrate post-quantum techniques, thereby providing researchers with novel avenues for exploration and development within the realm of quantum cryptography. This feature enhances the simulator's utility by accommodating evolving security challenges and fostering advancements in quantum cryptographic protocols.

In this study, a comprehensive simulation was conducted specifically focusing on the Key Generator Layer in conjunction with the default QKD KMS layer. The simulation results and analyses are presented in Tables 1 and 2, providing intricate details about node locations, layers, and the corresponding parameters employed in the QKDNetSim simulation.

Furthermore, Figure [10] visually encapsulates the statistics related to the generation and storage of key pairs in the QKD buffers. It also outlines the consumption patterns of key pairs by Alice and Bob's QKD systems.

Extending our analysis, Tables 3 and 4 furnish a detailed breakdown of node locations, layers, and associated parameters employed in the QKDNetSim simulation. Complementing this information, Figure [11] offers key insights into statistics concerning the generation and storage of key pairs in QKD buffers, as well as their consumption by Alice and Bob's QKD systems.

Figure [12] is dedicated to presenting statistics specifically related to the Key Consumer Layer ETSI 014, highlighting authentication type VMAC and encryption type One-Time Pad.

Moving forward, Tables 5 and 6 provide an exhaustive breakdown of node locations, layers, and associated parameters employed in the QKDNetSim simulation. Figure [13] visually represents key statistics pertaining to the generation, storage, and consumption of key pairs by Alice and Bob's QKD systems. Additionally, Figure [14] focuses on statistics for the Key Consumer Layer ETSI004 with authentication type VMAC and encryption type One-Time Pad.

Continuing the exploration, Tables 7 and 8 offer a comprehensive overview of node locations, layers, and relevant parameters utilized in the QKDNetSim simulation. Figure [15] complements this data by presenting key statistics related to the generation and storage of key pairs, along with their consumption by Alice and Bob's QKD systems. Furthermore, Figure [16] delves into statistics for the Key Consumer Layer ETSI 004, emphasizing authentication type VMAC and encryption type One-Time Pad. Finally, Figure [17] provides insights into statistics for the Key Consumer Layer ETSI 014, sharing authentication type VMAC and encryption type One-Time Pad.

\section{Conclusion}

The comparative analysis presented in \cite{14} underscores the vulnerabilities of the BB84 protocol to a spectrum of attacks, encompassing PNS, IRUD, BS, DoS, MAM, and IRA attacks. Consequently, ongoing research is actively exploring enhancements to the BB84 protocol. Integrating Quantum Key Distribution (QKD) protocols into the realm of the Internet of Things (IoT) and real-time networks is identified as a viable avenue, achievable through the implementation of a key generation layer utilizing the QKDNetSim simulation.

One of the paramount advantages of QKD lies in its ability to furnish a Secured Shared Key (SSK) via secure communication channels. It is imperative that the generated key demonstrates resilience against a myriad of potential attacks. As part of future research endeavors, there is a planned exploration of diverse QKD protocols, incorporating post-quantum cryptography and lattice-based cryptosystems. This proactive approach seeks to fortify the security landscape and ensure robust key generation in the face of evolving threat landscapes.
\pagebreak

\end{document}